\documentclass{article}
\usepackage{epsfig}
\usepackage{amssymb}

\newcommand{\be}{\begin{equation}}
\newcommand{\ee}{\end{equation}}
\newcommand{\ba}{\begin{eqnarray}}
\newcommand{\ea}{\end{eqnarray}}

\begin{document}

\title{{\bf The BCS-Bose Crossover Theory}}
\author{S.K. Adhikari,$^{a}$ M. de Llano,$^{b,c,e}$ F.J. Sevilla,$^{d,e}$
M.A. Sol\'{\i}s$^{d}$ \& J.J. Valencia$^{c,f}$ \\
%EndAName
$^{a}$Instituto de F\'{\i}sica Te\'{o}rica, UNESP- S\~{a}o Paulo State
University, 01405-900 S\~{a}o Paulo, SP, Brazil\\
$^{b}$Texas Center for Superconductivity, University of Houston, Houston, TX
77204, USA \and $^{c}$Instituto de Investigaciones en Materiales,
Universidad Nacional Aut\'{o}noma de M\'{e}xico  \and 04510 M\'{e}xico, DF,
Mexico \and $^{d}$Instituto de F\'{\i}sica, UNAM, 01000 M\'{e}xico, DF,
Mexico \and $^{e}$Consortium of the Americas for Interdisciplinary Science,
University of New Mexico \and Albuquerque, NM 87131, USA \and $^{f}$%
Universidad de la Ciudad de M\'{e}xico, San Lorenzo Tezonco, 09940 M\'{e}%
xico, DF, Mexico}
\maketitle

\begin{abstract}
We contrast {\it four} distinct versions of the BCS-Bose statistical
crossover theory according to the form assumed for the electron-number
equation that accompanies the BCS gap equation. The four versions correspond
to explicitly accounting for two-hole-(2h) as well as two-electron-(2e)
Cooper pairs (CPs), or both in equal proportions, or only either kind.\ This
follows from a recent generalization of the Bose-Einstein condensation
(GBEC) statistical theory that includes not boson-boson interactions but
rather 2e- and also (without loss of generality) 2h-CPs interacting with
unpaired electrons and holes in a single-band model that is easily converted
into a two-band model. The GBEC theory is essentially an extension of the
Friedberg-T.D. Lee 1989 BEC theory of superconductors that excludes 2h-CPs.
It can thus recover, when the numbers of 2h- and 2e-CPs in both BE-condensed
and noncondensed states are separately equal, the BCS gap equation for all
temperatures and couplings as well as the zero-temperature BCS
(rigorous-upper-bound) condensation energy for all couplings. But ignoring
either 2h- {\it or }2e-CPs it can do neither. In particular, only {\it half }%
the BCS condensation energy is obtained in the two crossover versions
ignoring either kind of CPs. We show how critical temperatures $T_{c}$ from
the original BCS-Bose crossover theory in 2D require unphysically large
couplings for the Cooper/BCS model interaction\ to differ significantly from
the $T_{c}$s of ordinary BCS theory (where{\it \ }the number equation is
substituted by the assumption that the chemical potential equals the Fermi
energy).

\noindent

PACS \# 74.20.-z; 74.20.Mn; 05.30.Fk; 05.30.Jp \newline

Key words: Bose-Einstein condensation statistical model; BCS-Bose crossover
theory.
\end{abstract}

\section{Introduction}

Boson-fermion (BF) statistical models of superconductivity (SC) as a
Bose-Einstein condensation (BEC) \cite{Ogg,Ginz} began to be seriously
studied in the mid-1950's \cite{BF}-\cite{Blatt}, pre-dating even the
BCS-Bogoliubov statistical theory \cite{bcs}-\cite{bts}. Although BCS theory
only contemplates the presence of ``Cooper correlations'' of single-particle
states, BF models \cite{BF}-\cite{Blatt}\cite{BF7a}-\cite{PhysicaC} posit
the existence of actual bosonic CPs. A drawback of early BF models is the
notorious absence of an electronic gap $\Delta (T)$, with $T$ the absolute
temperature. Perhaps the first BF model {\it with} a gap was introduced in
Ref. \cite{Eagles}. Somewhat later, the remarkable relation $\Delta (T) 
 \propto \sqrt{n_{0}(T)}$, with $n_{0}(T)$ the BEC condensate
number density of electron pairs, first seems to have appeared \cite{BF3}.
It resurfaced a year later in the BEC BF model in 3D of Friedberg and T.D.
Lee \cite{BF5,BF6} applied to cuprate superconductors. With just {\it one}
adjustable parameter (the ratio of perpendicular to $CuO_{2}$-plane boson
masses) this theory fitted \cite{BF6}\ quasi-2D cuprate $T_{c}/T_{F}$
empirical values \cite{Uemura89}\ rather well. The ratio turned out to be
66,560---just under the $10^{5}$ anisotropy ratio reported for $%
B_{2+x}Sr_{2-y}CuO_{2\pm \delta }$ \cite{Fiory} almost contemporaneously.

An extension of the work in\ Refs. \cite{BF5,BF6} is a generalized BEC
(GBEC) statistical single-band theory whereby a superconducting BCS
condensate was recently suggested \cite{BF7a}, and subsequently confirmed %
\cite{PLA2} (but only to lowest order in the BF coupling) to be precisely a
Bose-Einstein condensate (BEC) of equal numbers of bosonic two-electron (2e)
and two-hole (2h) Cooper pairs (CPs), at least inasmuch as the GBEC
reproduces the same BCS gap equation for all temperature and coupling as
well as the same $T=0$\ condensation energy found from BCS theory.\ The
holes that make up the 2h-CPs originate precisely the Fermi sea asssociated
with the $N$-electron system in the simple single-band model studied here.
One advantage of the single-band model is that it allows recovering, among
other theories, the BCS theory as a special case. The distinction (Ref. \cite%
{FW} pp. 70-72) between {\it single }particles and holes is, in a sense,
trivial. Not so for particle-pairs and hole-pairs, as will be seen shortly.

The BF coupling assumed appears in an interaction many-body Hamiltonian $%
H_{int}$\ which defines the GBEC theory. Added to $H_{int}$ is an
unperturbed Hamiltonian $H_{0}$\ describing a free {\it ternary} gas of
unpaired electrons plus 2e-CPs plus 2h-CPs. The noninteracting ternary gas
represents the normal state of the original, strongly-correlated
many-electron system under study, and is a viable candidate for a so-called
``non-Fermi-liquid.'' The new GBEC theory embodied in $H=H_{0}+H_{int}$\ is
in essence a {\it complete} BF (statistical) single-band model that,
however, admits departure \cite{PLA2} from the perfect 2e-/2h-CP symmetry
that constrains BCS theory by construction. It can be diagonalized via a
Bogoliubov canonical\ transformation{\it \ exactly} if one neglects nonzero
center-of-mass-momentum (CMM) CPs in $H_{int}$ as is done in BCS theory but 
{\it not }in $H_{0}$ which in BCS theory represents a pure electron gas. The
GBEC theory is {\it complete} only in the sense that 2h-CPs are not ignored.
It reduces to all the known statistical theories of superconductors (SCs),
including the BCS-Bose ``crossover'' picture in the four versions to be
distinguished below. Its practical impact is that it yields \cite%
{CMT02,deLlRev}\ robustly higher $T_{c}$'s than BCS theory without
abandoning electron-phonon dynamics when one departs from the perfect 50/50
symmetry of 2e-/2h-CPs in the condensate.

In the literature, electron-phonon dynamics have been widely mimicked by the
s-wave BCS/Cooper model interaction $V_{{\bf k,k^{\prime }}}$ \cite{bcs,Coo}%
.\ It is a nonzero negative constant $-V,$ if and only if single-particle
energies $\epsilon _{k},\epsilon _{k^{\prime }}$ lie within the energy
interval $[$max$\{0,\mu -\hbar \omega _{D}\},\mu +\hbar \omega _{D}]$ where $%
\mu $ is the electron chemical potential and $\omega _{D}$ the Debye
frequency. We employ this model interaction here. Other pairing symmetries
beyond pure s-wave can also be accommodated. Although it sheds considerable
light on different possible crossover schemes, our single-band picture where
single-electron and single-holes are assumed to have the same effective mass
is not as realistic in describing real materials as a {\it multiband} (say,
a valence-like band for holes and a conduction-like one for electrons)
theory where these two masses can differ, as will be discussed later.

A fundamental drawback of early \cite{BF}-\cite{Blatt} BF models, which took
2e CPs as analogous to diatomic molecules in a classical atom-molecule
binary gas mixture, is the cumbersome introduction of an electron energy gap 
$\Delta (T)$. ``Gapless'' models, however, are useful \cite{BF9,BF10}\ in
locating transition-temperature singularities if approached from {\it above}%
, i.e., from the {\it normal }state where $T>T_{c}$.

The ``ordinary'' CP problem \cite{Coo}\ for two distinct interfermion
interactions (the $\delta $-potential well \cite{PRB2000,PhysicaC} or the
Cooper/BCS model \cite{bcs,Coo} interactions) neglects the effect of 2h-CPs
treated on an equal footing with 2e\ [or, in general, two-particle (2p)]
CPs. On the other hand, Green's functions \cite{FW}\ can naturally deal with
hole propagation and thus accommodate both 2e- and 2h-CPs via, e.g., the
Bethe-Salpeter equation\ \cite{Honolulu,ANFdeLl}. In addition to the
generalized CP problem, a crucial result \cite{BF7a,PLA2}\ as already
mentioned is that the BCS condensate consists of equal numbers of 2e- and
2h-CPs. This was implicitly already suggested from the perfect symmetry\
about electron energy $\epsilon =\mu $ of the well-known Bogoliubov \cite%
{Bog} $v^{2}(\epsilon )$\ and $u^{2}(\epsilon )$ coefficients, with the tail
of $v^{2}(\epsilon )$ {\it above} $\epsilon =\mu $ representing 2e
correlations and that of $u^{2}(\epsilon )$ {\it below} $\epsilon =\mu $
refers to 2h correlations.

In this paper we show how: a) four versions of the BCS-Bose statistical
crossover theory can be obtained by ignoring either 2h- {\it or }2e-CPs or
by including both; c) for only two of the four versions can the precise BCS
gap equation for all temperatures $T$ be derived; c) crossover-picture $%
T_{c} $s, defined self-consistently by {\it both} the gap and fermion-number
equations, requires unphysically large couplings (at least for the
Cooper/BCS model interaction in 2D SCs)\ to differ significantly from the $%
T_{c}$ of ordinary BCS theory defined {\it without }the number equation
since here the chemical potential is assumed equal to the Fermi energy; and
d) the full $T=0$ BCS condensation energy follows from one crossover version
but only {\it half }of it from the two versions ignoring either kind of
CPs.\ The condensation energy is simply related to the ground-state energy
of the many-fermion system, which in the case of BCS is a rigorous upper
bound to the exact many-body value for the given Hamiltonian as BCS theory
starts from a variational wave function for the superconductor ground
state.\ These results, with the exception of (c) which does not apply, are
expected to hold also for neutral-fermion superfluids (SFs)---such as liquid 
$^{3}$He \cite{He3,Dobbs}, neutron matter and trapped ultra-cold fermion
atomic gases \cite{Holland}-\cite{jin2}---where the pair-forming two-fermion
interaction, of course, differs from the Cooper/BCS one for SCs.

\section{Generalized BEC Theory (GBEC)}

The GBEC theory is described in detail in Refs. \cite{BF7a}-\cite{deLlRev};
here we summarize its main equations.\ It applies in $d$ dimensions and is
defined by a Hamiltonian of the form $H=H_{0}+H_{int}$. The unperturbed
Hamiltonian $H_{0}$\ should ideally be, to quote Leggett \cite{Leggett06}
``an appropriate `zeroth-oder' starting point'' accounting for ``pairs of
electronic excitations with charge 2e that all have the same ground-state
wavefunction.''\ Thus, our $H_{0}$ corresponds to a non-Fermi-liquid
``normal'' state which, besides just fermions, is an ideal{\it \ }(i.e.,
noninteracting) ternary gas mixture of unpaired fermions {\it and} both
types of CPs namely, 2e and 2h, the latter introduced without loss of
generality. Specifically%
\begin{equation}
H_{0}=\sum\limits_{{\bf k}_{1},s_{1}}\epsilon _{{\bf k}_{{\bf 1}}}a_{{\bf k}%
_{1},s_{_{1}}}^{+}a_{{\bf k}_{1},s_{_{1}}}+\sum\limits_{{\bf K}}E_{+}(K)b_{%
{\bf K}}^{+}b_{{\bf K}}-\sum\limits_{{\bf K}}E_{-}(K)c_{{\bf K}}^{+}c_{{\bf K%
}}  \label{H0}
\end{equation}%
where ${\bf K\equiv k}_{{\bf 1}}+{\bf k}_{{\bf 2}}$ is the CMM wavevector of
the pair, while $\epsilon _{{\bf k}_{1}}\equiv \hbar ^{2}k_{1}^{2}/2m$ are
the single-electron, and $E_{\pm }(K)$\ the 2e-/2h-CP {\it phenomenological, 
}energies.\ Here $a_{{\bf k}_{1},s_{1}}^{+}$ ($a_{{\bf k}_{1},s_{1}}$) are
creation (annihilation) operators for fermions and similarly $b_{{\bf K}%
}^{+} $ ($b_{{\bf K}}$) and $c_{{\bf K}}^{+}$ ($c_{{\bf K}}$) for 2e- and
2h-CP bosons, respectively. These $b$ and $c$ operators depend only on ${\bf %
K}$\ and so are {\it distinct} from the BCS operators depending on both $%
{\bf K}$ {\it and} the relative ${\bf k}\equiv \frac{1}{2}{\bf (k}_{{\bf 1}}-%
{\bf k}_{{\bf 2}})$\ discussed in Ref. \cite{bcs} Eqs. (2.9) to (2.13) for
the particular case of ${\bf K=}$ $0$ and shown there {\it not} to satisfy
the ordinary Bose commutation relations.\ But because two pairs cannot
exactly overlap in real space without violating the Pauli Principle, they
are often considered ``hard-core bosons,'' albeit of hard-core radii $0^{+}$%
. For this reason, one can probably not expect to be able to construct the $%
b $ and $c$ operators directly from the $a$ operators in order to establish
that $b$ and $c$ obey Bose commutation relations precisely. Nonetheless,
these pairs stand for objects that can easily be seen to obey Bose-Einstein
statistics as, in the thermodynamic limit, an indefinitely large number of $%
{\bf k}$\ values correspond to a given ${\bf K}$ value defining an energy
level $E_{+}(K)$ or $E_{-}(K).$ This is all that is needed to ensure the BEC
(or macroscopic occupation of a given state that appears below a certain
fixed $T=T_{c}$) found \cite{BF7a}-\cite{deLlRev}\ numerically {\it a
posteriori }in the GBEC theory.\ Furthermore, being non-interacting (except
for the Pauli Principle restriction mentioned), CPs satisfy the
Ehrenfest-Oppenheimer \cite{Oppen}\ criteria for two clusters of charges to
conserve a specific kind of statistics, either Bose or Fermi. These assumed
properties are justified {\it a posteriori }when in the GBEC theory: a) the
BCS gap equation is recovered for equal numbers of both kinds of pairs, both
in the ${\bf K=}$ $0$ state and in all ${\bf K}\neq 0$ states taken
collectively, and in weak coupling, regardless of CP overlaps; and b) the
precise familiar BEC $T_{c}$ formula emerges \cite{BF7a}\ when i) 2h-CPs are
ignored, the Friedberg-T.D. Lee model \cite{BF5,BF6}\ equations are
recovered and ii) one switches off the BF interaction. The only difference
in the recovered BEC $T_{c}$ formula is that the boson number density now
depends on $T_{c}$, as expected in a boson-fermion mixture where populations
are $T$-dependent. Finally, we note that fermion scattering terms \cite%
{Nozieres85}\ are not included in \ref{H0} as they are not expected to be
substantial, say, in the BCS limit of high electron density where they would
be the most effective, which in turn is included in the GBEC model as a
special case.

Two-hole CPs in (\ref{H0}) are postulated to be {\it distinct }and{\it \
kinematically independent }from both the 2e-CPs and the unpaired electrons,
i.e., operators $a$, $b$ and $c$ are assumed to commute with each other.
This postulate is grounded on magnetic-flux-quantization measurements
establishing the presence of {\it pair }charge carriers in both conventional %
\cite{classical,classical2}\ as well as cuprate \cite{cuprates}\
superconductors, and on the fact that no experiment has yet been done, to
our knowledge \cite{Gough}, that{\large \ }distinguishes between electron
and hole CPs. The latter uncertainty further motivates a Hamiltonian such as
(\ref{H0}) with {\it both }kinds of CPs.

The interaction Hamiltonian $H_{int}$ in the expression $H=H_{0}+H_{int}$\
describes the formation and disintegration of CPs, respectively, from and
into unpaired electrons and holes. It is further simplified by dropping all $%
{\bf K}\neq 0$ terms. This is also done in BCS theory in its full{\it \ }%
Hamiltonian $H=H_{0}+H_{int}$,\ but {\it kept }in the GBEC theory in its
unperturbed $H_{0}$\ portion (\ref{H0}). The GBEC $H_{int}$ is made up of
four distinct BF interaction vertices each with two-fermion/one-boson
creation and/or annihilation operators. These vertices depict how unpaired
electrons (subindex +) [or holes (subindex $-$)] are involved in the
formation and disintegration\ of the 2e- (and 2h-) ${\bf K}=0$\ CPs in the $%
d $-dimensional system of size $L$, namely 
\begin{equation}
H_{int}=L^{-d/2}\sum\limits_{{\bf k}}f_{+}(k)\{a_{{\bf k},\uparrow }^{+}a_{-%
{\bf k},\downarrow }^{+}b_{{\bf 0}}+a_{-{\bf k},\downarrow }a_{{\bf k}%
,\uparrow }b_{{\bf 0}}^{+}\}+L^{-d/2}\sum\limits_{{\bf k}}f_{-}(k)\{a_{{\bf k%
},\uparrow }^{+}a_{-{\bf k},\downarrow }^{+}c_{{\bf 0}}^{+}+a_{-{\bf k}%
,\downarrow }a_{{\bf k},\uparrow }c_{{\bf 0}}\}  \label{Hint}
\end{equation}%
where ${\bf k}\equiv \frac{1}{2}{\bf (k}_{{\bf 1}}-{\bf k}_{{\bf 2}})$ is
again the relative wavevector of a CP. The interaction vertex form factors $%
f_{\pm }(k)$ in (\ref{Hint}) are\ essentially the Fourier transforms of the
2e- and 2h-CP intrinsic wavefunctions, respectively, in the relative
coordinate of the two fermions. The GBEC theory is thus reminiscent of the
Sommerfeld theory of the electron gas combined with the Debye picture of the
phonon gas which together give a binary mixture of noninteracting electrons
and phonons, a picture which describes low-$T$ specific heats in metals and
insulators. But to explain either resistance and superconductivity, they
must then be allowed to interact via the Fr\"{o}hlich electron-phonon
interaction \cite{Frohlich}\ of a form analogous to (\ref{Hint}) but without
hole terms. In contrast, the full BCS Hamiltonian $H_{0}^{BCS}+H_{int}^{BCS}$
consists of only the first (electron)\ term on the rhs of (\ref{H0}), namely 
\begin{equation}
H_{0}^{BCS}=\sum\limits_{{\bf k}_{1},s_{1}}\epsilon _{{\bf k}_{{\bf 1}}}a_{%
{\bf k}_{1},s_{_{1}}}^{+}a_{{\bf k}_{1},s_{_{1}}}  \label{H0BCS}
\end{equation}%
and%
\begin{equation}
H_{int}^{BCS} = \sum\limits_{{\bf k}_{1},{\bf l}_{1}}V_{{\bf k}_{1}%
{\bf ,l}_{1}}a_{{\bf k}_{1}\uparrow }^{+}a_{-{\bf k}_{1}\downarrow }^{+}a_{-%
{\bf l}_{1}\downarrow }a_{{\bf l}_{1}\uparrow .}  \label{HintBCS}
\end{equation}%
The BCS $H_{0}^{BCS}$ thus represents a Fermi liquid normal state.

The interaction vertex form factors $f_{\pm }(k)$ in (\ref{Hint}) are\
essentially the Fourier transforms of the 2e- and 2h-CP intrinsic
wavefunctions, respectively, in the relative coordinate between the paired
fermions of the CP. In order to eventually recover BCS theory, in Refs. \cite%
{BF7a}-\cite{deLlRev} the corresponding energy form factors were picked as%
\begin{equation}
f_{+}(\epsilon )=\left\{ 
\begin{array}{cc}
f & \quad \mbox{for}\,\,E_{f}<\epsilon <E_{f}+\delta \varepsilon \quad  \\ 
0 & \mbox{otherwise,}%
\end{array}%
\right.   \label{f+}
\end{equation}%
\begin{equation}
f_{-}(\epsilon )=\left\{ 
\begin{array}{cc}
f & \quad \mbox{for}\,\,E_{f}-\delta \varepsilon <\epsilon <E_{f}\quad  \\ 
0 & \mbox{otherwise.}%
\end{array}%
\right.   \label{f-}
\end{equation}%
This is after one introduces the quantities $E_{f}$ and $\delta \varepsilon $
as {\it new} phenomenological dynamical energy parameters (in addition to
the positive BF vertex coupling parameter $f$) that replace the previous
phenomenological CP energy parameters $E_{\pm }(0)$, through the definitions 
\begin{equation}
E_{f}\equiv {\textstyle\frac{1}{4}}[E_{+}(0)+E_{-}(0)]\ \ \ \mbox{and}\ \ \
\delta \varepsilon \,\equiv \,{\textstyle\frac{1}{2}}[E_{+}(0)-E_{-}(0)]\geq
0  \label{27}
\end{equation}%
where $E_{+}(0)$ and $E_{-}(0)$ are the (empirically {\it un}known) zero-CMM
energies of the 2e- and 2h-CPs, respectively. Note that $2E_{f}$ lies midway
between $E_{+}(0)$ and $E_{-}(0).$ Alternately, instead of (\ref{27}) one
can write the two relations 
\begin{equation}
E_{\pm }(0)=2E_{f}\pm \delta \varepsilon .  \label{Eplusminus(0)}
\end{equation}%
The quantity\ $E_{f}$ serves as a convenient energy scale; it is not to be
confused with the Fermi energy $E_{F}={\textstyle\frac{1}{2}}%
mv_{F}^{2}\equiv k_{B}T_{F}$ where $T_{F}$\ is the Fermi temperature. The
Fermi energy $E_{F}$ equals $\pi \hbar ^{2}n/m$ in 2D and $(\hbar
^{2}/2m)(3\pi ^{2}n)^{2/3}$ in 3D, with $n\equiv N/L^{d}$ the total
number-density of charge-carrier mobile electrons, while $E_{f}$ is of the
same form but with $n$ replaced by, say, $n_{f}$, which in turn serves as
convenient electron-density scale. The quantities $E_{f}$ and $E_{F}$
coincide {\it only }when perfect 2e/2h-CP symmetry holds, i.e., when $n=n_{f}
$.

The grand potential $\Omega $\ for the full Hamiltonian $H=H_{0}+H_{int}$
given by (\ref{H0}) and (\ref{Hint}) is then constructed via (Ref. \cite{FW}
Eq. 4.14) the definition%
\begin{equation}
\Omega (T,L^{d},\mu ,N_{0},M_{0}){\ \ }=-k_{B}T\ln \left[ \mbox{Tr}e^{-\beta
(H-\mu \hat{N})}\right]   \label{28}
\end{equation}%
where ``Tr'' stands for ``trace'' and $\beta \equiv 1/k_{B}T$ with $T$ the
absolute temperature. It is related to the system pressure $P$, internal
energy $E$ and entropy $S$ by $\Omega =-PL^{d}=F-\mu N=E-TS-\mu N$, where $F$
is the Helmholtz free energy. Following the Bogoliubov prescription \cite%
{Bog47}, one sets $b_{{\bf 0}}^{+},$ $b_{{\bf 0}}$ equal to $\sqrt{N_{0}}$
and $c_{{\bf 0}}^{+}$, $c_{{\bf 0}}$ equal to $\sqrt{M_{0}}$ in (\ref{Hint}%
), where $N_{0}$ is the $T$-dependent number of zero-CMM 2e-CPs and $M_{0}$
likewise for 2h-CPs. This allows {\it exact }diagonalization for any
coupling, through a Bogoliubov transformation of the $a^{+},a$ fermion
operators, giving \cite{Casas04} after some algebra%
\begin{eqnarray}
\Omega (T,L^{d},\mu ,N_{0},M_{0})/L^{d} &=&\int_{0}^{\infty }d\epsilon
N(\epsilon )[\epsilon -\mu -E(\epsilon )]-2k_{B}T\int_{0}^{\infty }d\epsilon
N(\epsilon )\ln \{1+\exp [-\beta E(\epsilon )]\}  \nonumber \\
&&+[E_{+}(0)-2\mu ]n_{0}+k_{B}T\int_{0^{+}}^{\infty }d\varepsilon
M(\varepsilon )\ln \{1-\exp [-\beta \{E_{+}(0)+\varepsilon -2\mu \}]\} 
\nonumber \\
&&+[2\mu -E_{-}(0)]m_{0}+k_{B}T\int_{0^{+}}^{\infty }d\varepsilon
M(\varepsilon )\ln \{1-\exp [-\beta \{{2\mu -E}_{-}(0)+\varepsilon \}]\}.
\label{omega}
\end{eqnarray}%
Here $N(\epsilon )$ and $M(\varepsilon )$ are respectively the electronic
and bosonic density of states, while 
\begin{equation}
E(\epsilon )\equiv \sqrt{(\epsilon -\mu )^{2}+\Delta ^{2}(\epsilon )}\equiv 
\sqrt{(\epsilon -\mu )^{2}+n_{0}f_{+}^{2}(\epsilon )+m_{0}f_{-}^{2}(\epsilon
)}  \label{E}
\end{equation}%
since $\Delta (\epsilon )\equiv \sqrt{n_{0}}f_{+}(\epsilon )+\sqrt{m_{0}}%
f_{-}(\epsilon )$ and $f_{+}(\epsilon )f_{+}(\epsilon )\equiv 0$\ from (\ref%
{f+}) and (\ref{f-}), with $n_{0}(T)\equiv N_{0}(T)/L^{d}$ and $%
m_{0}(T)\equiv M_{0}(T)/L^{d}$ being the 2e-CP and 2h-CP number densities,
respectively, of BE-condensed (i.e., with $K=0$) bosons.

Minimizing $F$\ with respect to $N_{0}$ and $M_{0}$, while simultaneously
fixing the total number $N$ of electrons by introducing the electron
chemical potential $\mu $ in the usual way, namely\ 
\begin{equation}
\frac{\partial F}{\partial N_{0}}{\ \ }={\ \ }0,\,{\ \ \ \ \ \ }\frac{%
\partial F}{\partial M_{0}}{\ \ \ }={\ \ }0,\qquad \mbox{and}\qquad \frac{%
\partial \Omega }{\partial \mu }{\ \ }={\ \ }-N  \label{36}
\end{equation}%
ensures an {\it equilibrium thermodynamic state} of the system with volume $%
L^{d}$ at temperature $T$ and chemical potential $\mu $. Evidently, $N$
includes both paired and unpaired CP electrons. Some algebra then leads \cite%
{Casas04}\ to the three coupled integral Eqs. (7)-(9) of Ref. \cite{BF7a}
which, since from (\ref{f+}) and (\ref{f-}) one has that $f_{+}(\epsilon
)f_{-}(\epsilon )\equiv 0$, can be simplified to the two {\it ``gap-like
equations''}%
\begin{equation}
\lbrack 2E_{f}+\delta \varepsilon -2\mu (T)]=%
%TCIMACRO{\U{bd}}%
%BeginExpansion
{\frac12}%
%EndExpansion
f^{2}\int\limits_{E_{f}}^{E_{f}+\delta \varepsilon }d\epsilon N(\epsilon )%
\frac{\tanh 
%TCIMACRO{\U{bd}}%
%BeginExpansion
{\frac12}%
%EndExpansion
\beta \sqrt{\lbrack \epsilon -\mu (T)]^{2}+f^{2}n_{0}(T)}}{\sqrt{[\epsilon
-\mu (T)]^{2}+f^{2}n_{0}(T)}}  \label{gaplike1}
\end{equation}%
\begin{equation}
\lbrack 2\mu (T)-2E_{f}+\delta \varepsilon ]=%
%TCIMACRO{\U{bd}}%
%BeginExpansion
{\frac12}%
%EndExpansion
f^{2}\int\limits_{E_{f}-\delta \varepsilon }^{E_{f}}d\epsilon N(\epsilon )%
\frac{\tanh 
%TCIMACRO{\U{bd}}%
%BeginExpansion
{\frac12}%
%EndExpansion
\beta \sqrt{\lbrack \epsilon -\mu (T)]^{2}+f^{2}m_{0}(T)}}{\sqrt{[\epsilon
-\mu (T)]^{2}+f^{2}m_{0}(T)}}  \label{gaplike2}
\end{equation}%
and a single {\it ``number equation'' }(that guarantees charge conservation)%
\begin{equation}
2n_{B}(T)-2m_{B}(T)+n_{f}(T)=n.  \label{number}
\end{equation}%
where%
\begin{equation}
n_{f}(T)\equiv \int\limits_{0}^{\;\infty }d\epsilon N(\epsilon )[1-\frac{%
\epsilon -\mu }{E(\epsilon )}\tanh 
%TCIMACRO{\U{bd}}%
%BeginExpansion
{\frac12}%
%EndExpansion
\beta E(\epsilon )]  \label{nfEQ}
\end{equation}%
is clearly the number of unpaired electrons. This is identical with $2\sum_{%
{\bf k}}v_{k}^{2}(T)$\ with $v_{k}^{2}(T)$ the well-known $T${\it -dependent}%
\ Bogoliubov $v^{2}$-coefficient. In (\ref{number}) $n\equiv N/L^{d}$ is the
number density of electrons while\ $n_{B}(T)$ and $m_{B}(T)$ are,
respectively, the number densities of 2e- and 2h-CPs in {\it all }bosonic
states (both $K=0$ as well as $K>0$). The ``complete'' number equation (\ref%
{number}) can be rewritten more explicitly as%
\begin{equation}
2n_{0}(T)+2n_{B+}(T)-2m_{0}(T)-2m_{B+}(T)+n_{f}(T)=n\ \ \ \ \ \ \ \ \ %
\mbox{\bf crossover version A}  \label{57'}
\end{equation}%
where $n_{B}(T)$ is \cite{deLlRev}%
\begin{equation}
n_{B}(T)\equiv n_{0}(T)+n_{B+}(T);{\ \ \ \ }n_{B+}(T)\equiv
\int\limits_{0+}^{\infty }d\varepsilon M(\varepsilon )[\exp \beta
\{E_{+}(0)+\varepsilon -2\mu \}-1]^{-1}  \label{nB}
\end{equation}%
and similarly for $m_{B}(T)$ which is%
\begin{equation}
m_{B}(T)\equiv m_{0}(T)+m_{B+}(T);{\ \ \ \ }m_{B+}(T)\equiv
\int\limits_{0+}^{\infty }d\varepsilon M(\varepsilon )[\exp \beta \{{2\mu }{-%
}E_{-}(0)+\varepsilon \}-1]^{-1}.  \label{mB}
\end{equation}%
Clearly, $m_{B+}(T)$ are precisely the number of ``pre-formed'' $K>0$\
2h-CPs, and $n_{B+}(T)$\ that of 2e-CPs. These CPs are non-condensed in
contrast with the $K=0$ CPs which are BE condensed. Evaluating the integrals
requires knowing the bosonic density-of-states $M(\varepsilon )$ of CPs of
energy $\varepsilon ,$ which in turn requires knowing the dispersion
relation $\varepsilon $ vs. $K$, e.g., as has been determined via the
Bethe-Salpeter equation in the ladder approximation in 3D \cite{Honolulu}\
and in 2D \cite{ANFdeLl}.

Self-consistent (at worst, numerical) solution of the three coupled integral
equations{\bf \ }(\ref{gaplike1}), (\ref{gaplike2}) and (\ref{57'}) then
yields the three thermodynamic variables of the GBEC theory 
\begin{equation}
\ \ n_{0}(T,n,\mu ),\;\ \ \ m_{0}(T,n,\mu ),\qquad \mbox{and}\qquad \mu
(T,n).  \label{43}
\end{equation}%
Figure 1 displays the three BE condensed phases---labeled $s+$, $s-$ and $ss$%
---along with the normal phase $n$, that emerge \cite{PLA2} from the GBEC
theory. Phase $s+$ stands for a pure 2e-CP BE condensate, $s-$ for a pure
2h-CP such condensate and $ss$ denotes a mixed phase. Only the two pure
phases were found \cite{PLA2} to display $T_{c}$ values higher than the
corresponding BCS value, while the mixed phase occurs below this value. 

%FIGURE 1
\begin{figure}[tbh]
\centerline{\epsfig{file=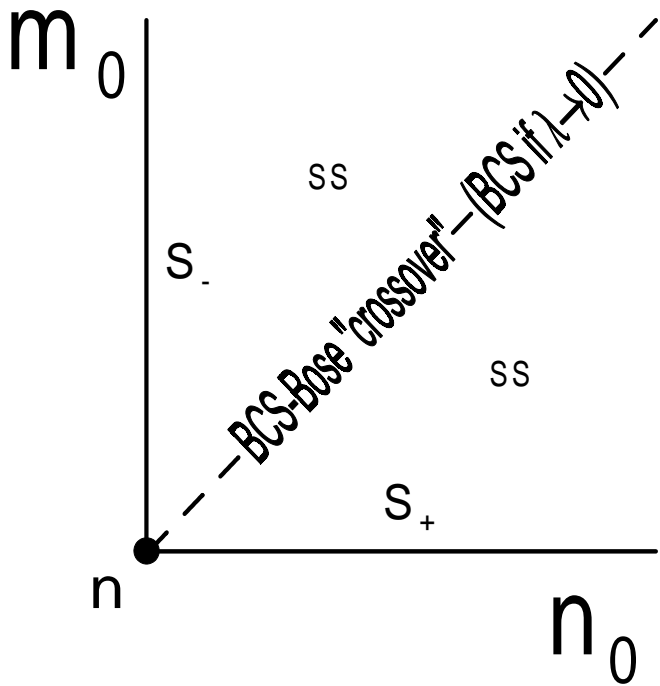,height=3.8in,width=3.6in}}
\caption{ Illustration in the $n_{0}$-$m_{0}$ plane of three GBEC theory
condensed phases (the pure 2e-CP $s+$ and pure 2h-CP $s-$ BE condensate
phases and a mixed phase $ss$) along with the normal (ternary BF
non-Fermi-liquid) phase $n$ that corresponds to the origin at $n_{0}=0=m_{0}.
$}
\end{figure}

For the two pure phases one can, in principle, shift from the single-band
model implied so far to a {\it two}-band model by allowing the particle (e)
masses to differ from hole (h) masses; this can be done by introducing two
different Fermi energies $E_{F}^{e}$ and $E_{F}^{h}$ that differ precisely
by these two masses.

The GBEC theory contains \cite{deLlRev}\ the key equations of all {\it five }%
distinct statistical theories as special cases. These range from ordinary
BCS to ordinary BEC theories, which are thereby completely unified by the
GBEC theory. Perfect 2e/2h CP symmetry signifies equal numbers of 2e- and
2h-CPs, more specifically, $n_{B}(T)=m_{B}(T)$ {\it as well as }$%
n_{0}(T)=m_{0}(T).$ This implies that $n_{B+}(T)=m_{B+}(T)$ for all $T,$
meaning that the exponents in (\ref{nB}) and (\ref{mB}) coincide so that
with (\ref{Eplusminus(0)}) this makes $E_{f}=\mu $. The GBEC theory then
reduces to the gap and number equations [viz., in 2D for $T=T_{c}$\ both (%
\ref{2D-gap-1}) and (\ref{2D-Num}) below]\ of the original \cite{Friedel} 
{\it BCS-Bose crossover picture} with the Cooper/BCS model interaction---if
its parameters $V$ and $\hbar \omega _{D}$ are identified with the BF
interaction GBEC Hamiltonian $H_{int}$ parameters $f^{2}/2\delta \varepsilon 
$ and $\delta \varepsilon $, respectively.\ This one-to-one correspondence
between $H_{int}$ and $H_{int}^{BCS}$\ defined in (\ref{Hint}) and (\ref%
{HintBCS}) justifies the particular choice of form factors (\ref{f+}) and (%
\ref{f-}) for the BF interaction.\ The original crossover picture for
unknowns $\Delta (T)$ and $\mu (T)$ is now supplemented by the central
relation 
\begin{equation}
\Delta (T)=f\sqrt{n_{0}(T)}=f\sqrt{m_{0}(T)}.  \label{keyrelation}
\end{equation}%
All three functions $\Delta (T),$ $n_{0}(T)$\ and $m_{0}(T)$\ have the
familiar ``half-bell-shaped'' forms. Namely, they are zero above a certain
critical temperature $T_{c}$, and rise monotonically upon cooling (lowering $%
T$) to maximum values $\Delta (0),$\ $n_{0}(0)$\ and $m_{0}(0)$\ at\ $T=0.$
The energy gap $\Delta (T)$\ is the order parameter describing the
superconducting (or superfluid) condensed state,\ while $n_{0}(T)$\ and\ $%
m_{0}(T)$\ are the BEC order parameters depicting the macroscopic occupation
that occurs below $T_{c}$ in a BE condensate. This $\Delta (T)$\ is
precisely the BCS energy gap if the boson-fermion coupling $f$ is made to
correspond to $\sqrt{2V\hbar \omega _{D}}$ within the GBEC formalism.
Evidently, the BCS and BEC $T_{c}$s are the same. Writing (\ref{keyrelation}%
) for $T=0$ and dividing this into (\ref{keyrelation}) gives the much
simpler $f$-independent relation involving order parameters {\it normalized}
to unity in the interval [$0,1$] 
\begin{eqnarray}
\Delta (T)/\Delta (0) &=&\sqrt{n_{0}(T)/n_{0}(0)}\ =\sqrt{m_{0}(T)/m_{0}(0)}{%
\ \ }\quad \smash {\mathop{\relbar\joinrel\longrightarrow}\limits_{T \to 0}}%
\quad 1  \nonumber \\
&&{\ \ }\quad \quad \quad \quad \quad \quad \quad \quad \quad \quad \quad
\quad \quad \quad \quad \quad \smash
{\mathop{\relbar\joinrel\longrightarrow}\limits_{T \geq T_c}}\quad 0.
\label{universal}
\end{eqnarray}%
The first equality, apparently first obtained in Ref. \cite{BF3}, connects
in a simple way the two heretofore unrelated ``half-bell-shaped'' order
parameters of the BCS and the BEC theories.\ The second equality \cite%
{BF7a,PLA2}\ implies that a BCS{\Huge \ }condensate is precisely a BE
condensate of equal numbers of 2e- and 2h-CPs. Since (\ref{universal}) is 
{\it independent }of the particular two-fermion dynamics of the problem, it
can be expected to hold for either SCs and SFs.

\section{Gap equation}

The standard procedure in all SC and SF theories of many-fermion systems is
to {\it ignore }dealing explicitly with{\it \ }2h-CPs altogether. Neglecting
in (\ref{omega})\ all terms containing $m_{0}(T)$, $E_{-}(0)$ and $%
f_{-}(\epsilon )$ leaves an $\Omega (T,L^{d},\mu ,N_{0})$ defining a {\it %
binary}, instead of ternary,{\it \ }BF model.\ Minimizing the associated
Helmholtz free energy $F(T,L^{d},\mu ,N_{0})=\Omega (T,L^{d},\mu ,N_{0})+\mu
N$ over $N_{0}$\ (for fixed total electron number $N$) requires that $%
\partial F/\partial N_{0}=0=\partial F/\partial n_{0}$, which becomes 
\[
%\begin{equation*}
\int_{0}^{\infty }d\epsilon N(\epsilon )\left[ -1+\frac{2\exp \{-\beta
E(\epsilon )\}}{1+\exp \{-\beta E(\epsilon )\}}\right] \frac{dE(\epsilon )}{%
dn_{0}}+\left[ E_{+}(0)-2\mu \right] =0%\end{equation*}
\]%
or 
\begin{equation}
2\left[ 2E_{f}+\delta \varepsilon -2\mu \right] =f^{2}\int_{E_{f}}^{E_{f}+%
\delta \varepsilon }d\epsilon N(\epsilon )\frac{1}{E(\epsilon )}\tanh 
%TCIMACRO{\U{bd}}%
%BeginExpansion
{\frac12}%
%EndExpansion
\beta E(\epsilon ).  \label{2eGaplikeEq}
\end{equation}%
Using (\ref{Eplusminus(0)}) yields precisely the BCS gap equation for all $T$%
, Eq. (3.27) of Ref. \cite{bcs}, {\it provided one picks}$\ E_{f}=\mu $,
namely 
\begin{equation}
1=\lambda \int_{0}^{\hbar \omega _{D}}d\xi \frac{1}{\sqrt{\xi ^{2}+\Delta
^{2}(T)}}\tanh 
%TCIMACRO{\U{bd}}%
%BeginExpansion
{\frac12}%
%EndExpansion
\beta \sqrt{\xi ^{2}+\Delta ^{2}(T)}  \label{gap+}
\end{equation}%
where $\xi \equiv \epsilon -\mu $, since $\lambda \equiv
N(E_{F})V=f^{2}N(E_{F})/2\delta \varepsilon $ while $\delta \varepsilon
=\hbar \omega _{D}$ [see relation between $V$ and $f$ stated just above (\ref%
{keyrelation})], and provided $N(\epsilon )$ can be taken outside the
integral sign in (\ref{2eGaplikeEq}). This last operation is exact in 2D
when $N(\epsilon )$ is independent of $\epsilon $ and is otherwise a good
approximation if $\hbar \omega _{D}\ll \mu .$ 

However, the choice $E_{f}=\mu $ {\it cannot} be justified, to our
knowledge, without assuming within the GBEC that $n_{B}(T)=m_{B}(T)$ {\it as
well as }$n_{0}(T)=m_{0}(T),$ i.e., by explicitly recognizing the existence
of 2h-CPs along with 2e-CPs and taking them in equal or 50-50 proportions.

\section{Number equation}

Besides the {\it normal }phase consisting of the ideal BF ternary gas
described by $H_{0}$, three different stable BEC phases emerge \cite{PLA2}
when solving all three equations (\ref{gaplike1}) to (\ref{gaplike2}) or (%
\ref{57'}): two pure phases, a pure 2e-CP BEC and a pure 2h-CP BEC, as well
as a mixed phase consisting of both types of BECs in varying proportions.
For a half-and-half mixed phase, i.e., $n_{0}(T)=m_{0}(T)$ and $%
n_{B+}(T)=m_{B+}(T),$ all the boson number-density terms in (\ref{57'})\
cancel and the BCS number equation 
\begin{equation}
n=n_{f}(T)\ \ \ \ \ \ \ \ \ 
\mbox{\bf crossover version B (special case of
A)}  \label{BCSnumbereqn}
\end{equation}%
is recovered, with $n_{f}(T)$ defined by (\ref{nfEQ}).\ Crossover version B
does not{\it \ explicitly }neglect either kind of CP, nor does it draw a
distinction between them. and is the version applied below in 2D to obtain (%
\ref{2D-gap-1}) to (\ref{mu-2D}) and the results of Fig. 2. This is the
original crossover version first presented in 1967 in Ref. \cite{Friedel}.

If 2h-CPs are ignored altogether, the companion number equation follows from
the last equation of (\ref{36}) as 
\begin{equation}
n=n_{f}(T)+2n_{B}(T)\ \ \ \ \ \ \ \ \ \mbox{\bf crossover version C}
\label{neq2e}
\end{equation}%
where $n_{f}(T)$ is interpreted as the number density of unpaired but
BCS-correlated electrons and is given by (\ref{nfEQ}). In perhaps the first
attempt to discuss \cite{Eagles}\ BEC in 1969 within the BCS-Bose crossover
picture, Eagles \cite{Eaglesprivcomm}\ imposed (\ref{neq2e}){\it \ }to
accompany the gap equation in what was perhaps the first BF model {\it with }%
a gap. This {\it differs}\ from the much simpler number equation of
crossover version B, which gave (\ref{2D-Num}) below as a special case for $%
T=T_{c}$ when $\Delta (T_{c})=0$ is substituted into (\ref{BCSnumbereqn})
and (\ref{nfEQ}) if one uses the identity $1-\tanh (x/2)\equiv 2/(\exp x+1).$
It is this version that corresponds to the Friedberg-T.D. Lee model \cite%
{BF5,BF6}.

Similarly, ignoring 2e-CPs and keeping only 2h-CPs leads to $\Omega
(T,L^{d},\mu ,M_{0})$ from which to minimize $F(T,L^{d},\mu ,M_{0})$\ over%
{\LARGE \ }$M_{0}$ requires that one set $\partial F/\partial
M_{0}=0=\partial F/\partial m_{0}$. Noting that $E(\xi )\equiv E(-\xi ),$
this also leads to the gap equation (\ref{gap+}) provided, {\it again}, one
picks$\ E_{f}=\mu $, but now with the companion number equation 
\begin{equation}
n=n_{f}(T)-2m_{B}(T) \ \ \ \ \ \ \ \ \ \mbox{\bf crossover version D}
\label{neq2h}
\end{equation}%
instead of (\ref{neq2e}) but with the same $n_{f}(T)$ as in (\ref{nfEQ}).

\section{BCS-Bose crossover $T_{c}$ compared with BCS $T_{c}$ in 2D}

The original crossover theory \cite{Friedel}\ is defined by two simultaneous
coupled equations, the BCS gap and number equations, {\it without }the BCS
assumption that the chemical potential $\mu $ equals the Fermi energy $E_{F}$%
. For subsequent extensions of the original version, see reviews in Refs. %
\cite{BCS-Bose,LevinPhysReps}. The critical temperature $T_{c}$ is defined
by $\Delta (T_{c})=0$, and is to be determined self-consistently with $\mu
(T_{c})$ by solving both gap and number equations. Because of its interest
in quasi-2D cuprate superconductors \cite{Brandow}, in this section we
concentrate on 2D only. For the Cooper/BCS model interaction, if $\lambda
\equiv N(E_{F})V$ where $N(E_{F})=m/2\pi \hbar ^{2}$, the two crossover
equations to be solved self-consistently reduce to 
\begin{equation}
1=\lambda \int_{0}^{\hbar \omega _{D}/2k_{B}T_{c}}dx\frac{\tanh x}{x}\ \ \ (%
\mbox{if}\ \ \mu >\hbar \omega _{D});\ \ \ \ \ 1=\lambda \int_{-\mu
(T_{c})/2k_{B}T_{c}}^{\hbar \omega _{D}/2k_{B}T_{c}}dx\frac{\tanh x}{2x}\ \
\ (\mbox{if}\ \ \mu <\hbar \omega _{D})  \label{2D-gap-1}
\end{equation}%
\begin{equation}
\int_{0}^{\infty }\frac{d\epsilon }{\exp \{[\epsilon -\mu
(T_{c})]/2k_{B}T_{c}\}+1}=E_{F}.  \label{2D-Num}
\end{equation}%
The last integral can be done analytically and leaves the explicit
expression 
\begin{equation}
\mu (T_{c})=k_{B}T_{c}\ln (e^{E_{F}/k_{B}T_{c}}-1)  \label{mu-2D}
\end{equation}%
which is then eliminated symbolically from (\ref{2D-gap-1}) to render $T_{c}$
as an implicit function of $\lambda $ alone. Using $\hbar \omega
_{D}/E_{F}=0.05$ as a typical value for cuprates, increasing $\lambda $
makes $\mu (T_{c})$ decrease from its weak-coupling (where $T_{c}\rightarrow
0$) value of $E_{F}$ down to $\hbar \omega _{D}$ when $\lambda \simeq 56$,
an unphysically large value as it well exceeds the Migdal ionic-lattice
stability upper limit \cite{Migdal}\ of $1/2$, although \cite{Carbotte}\
``there is no universally accepted, simple, and quantitative stability
criterion.''

Fig. 2 displays $T_{c}$ (in units of $T_{F}$) as function of $\lambda $.
Note that room-temperature SCs (RTSC) are predicted by BCS-Bose crossover
theory but only for $\lambda $ values definitely larger than about $10$ that
are still too unphysical.

\begin{figure}[tbh]
\begin{center}
\centerline{\epsfig{file=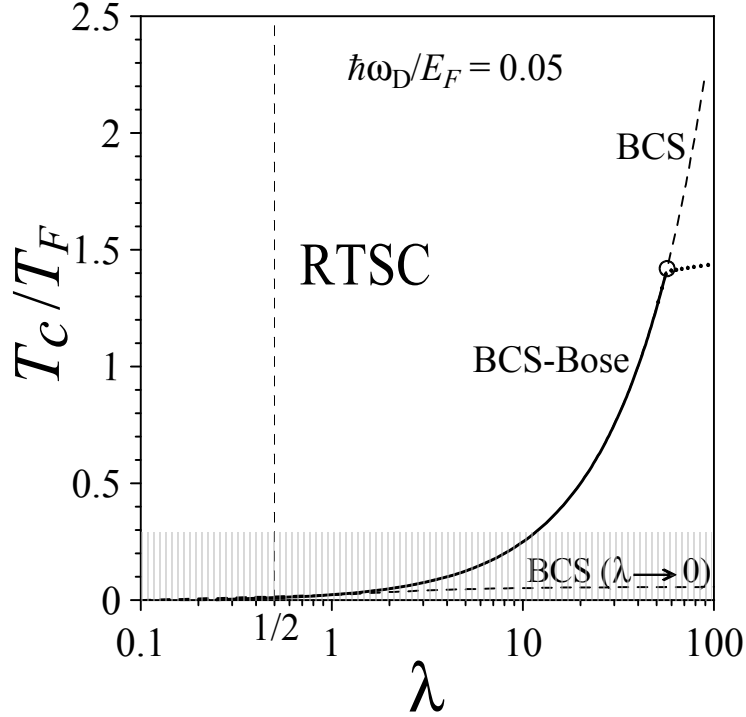,height=4in,width=4.0in}}
\end{center}
\par
\vspace{-1.0cm}
\caption{Critical SC temperatures $T_{c}$ in units of $T_{F}$ for the
BCS-Bose crossover theory (full curve) in 2D compared with the BCS value
from the exact implicit $T_{c}$ equation (see, e.g., Ref. \protect\cite{FW},
p. 447) $1$ $=\protect\lambda \protect\int_{0}^{\hbar \protect\omega %
_{D}/2k_{B}T_{c}}dxx^{-1}\tanh x$ (upper dashed curve) valid for {\it any }$%
d>0$ in any coupling $\protect\lambda $, and its weak-coupling explicit
solution $T_{c}\simeq 1.134\hbar \protect\omega _{D}\exp (-1/\protect\lambda %
)$ (lower dashed curve). The dot-dashed ``appendage'' signals a breakdown in
the BCS/Cooper interaction model when $\protect\mu (T_{c})$ turns negative,
as the Fermi surface at $\protect\mu $ then washes out and the interaction
model becomes meaningless. The value of $\protect\lambda =1/2$\ marked is
the maximum possible value allowed just short of lattice instability in 3D
for this interaction model, at least by one criterion \protect\cite{Migdal}.
Unshaded region refers to room-temperature superconductivity (RTSC) for SCs
with $T_{F} \lesssim 10^{3}$ K. }
\end{figure}

\section{Condensation energy}

The $T=0$ condensation energy per unit volume according to the GBEC theory,
given (\ref{omega}),\ is 
\begin{equation}
\frac{E_{s}-E_{n}}{L^{d}}=\frac{\Omega _{s}(T=0)-\Omega _{n}(T=0)}{L^{d}}
\label{Ec}
\end{equation}%
since for any $T$\ the Helmholtz free energy $F\equiv E-TS=\Omega +\mu N$,
with $S$ the entropy, and $\mu $ is the same for either superconducting $s$
or normal $n$ phases with internal energies $E_{s}$ and $E_{n}$,
respectively. In the {\it normal} phase $n_{0}(T)=0$, $m_{0}(T)=0$ so that $%
\Delta (T)=0$ for all $T\geq 0$, so that (\ref{omega}) reduces to 
\begin{equation}
\frac{\Omega _{n}(T=0)}{L^{d}}=\int_{0}^{\infty }d\epsilon N(\epsilon
)(\epsilon -\mu -|\epsilon -\mu |)=2\int_{0}^{\mu }d\epsilon N(\epsilon
)(\epsilon -\mu )=2\int_{-\mu }^{0}d\xi N(\xi )\xi .  \label{omega_n}
\end{equation}%
For the superconducting phase, and when $n_{0}(T)=m_{0}(T)$ and $%
n_{B}(T)=m_{B}(T)$ hold, i.e., crossover scenario B, one deduces from (\ref%
{Eplusminus(0)}) and (\ref{omega}) that $\mu =E_{f}$. Putting $\Delta (T=0)$ 
$\equiv \Delta $ in (\ref{omega}) as well as $\delta \varepsilon \equiv
\hbar \omega _{D}$, while using (\ref{Eplusminus(0)}), gives%
\begin{eqnarray}
\frac{\Omega _{s}(T=0)}{L^{d}} &=&2\hbar \omega _{D}n_{0}(0)+\int_{-\mu
}^{\infty }d\xi N(\xi )\left( \xi -\sqrt{\xi ^{2}+\Delta ^{2}}\right) 
\nonumber \\
&=&2\hbar \omega _{D}n_{0}(0)+2\int_{-\mu }^{-\hbar \omega _{D}}d\xi N(\xi
)\xi -2\int_{0}^{\hbar \omega _{D}}d\xi N(\xi )\sqrt{\xi ^{2}+\Delta ^{2}}.
\label{omega_s}
\end{eqnarray}%
The first factor of $2$\ in the last line comes precisely from the condition 
$n_{0}(T)=m_{0}(T)$ while the last two factors of $2$ arise from the
condition that according to (\ref{f+}) and (\ref{f-}) the magnitudes of $%
f_{+}(\epsilon )$ and $f_{-}(\epsilon )$ are the {\it same} and equal $f$.
Subtracting (\ref{omega_n}) from (\ref{omega_s}) and putting $N(\xi )\cong
N(0)$, the density of electronic states at the Fermi surface [designated
before as $N(E_{F})$]\ yields%
\[
\frac{E_{s}-E_{n}}{L^{d}}=2\hbar \omega _{D}n_{0}(0)+2N(0)\int_{0}^{\hbar
\omega _{D}}d\xi \left( \xi -\sqrt{\xi ^{2}+\Delta ^{2}}\right) 
\]%
\begin{equation}
=2\hbar \omega _{D}n_{0}(0)+N(0)\left[ (\hbar \omega _{D})^{2}-\hbar \omega
_{D}\sqrt{(\hbar \omega _{D})^{2}+\Delta ^{2}}+\Delta ^{2}\ln \frac{\Delta }{%
\hbar \omega _{D}+\sqrt{(\hbar \omega _{D})^{2}+\Delta ^{2}}}\right] \hspace{%
1cm}\hbox{(GBEC)}  \label{ocuarto}
\end{equation}%
exactly, by standard integrations. Using the expression that follows from (%
\ref{gap+}) for $T=0$ gives Eq. (2.40) of Ref. \cite{bcs}, namely%
\begin{equation}
\Delta =\frac{\hbar \omega _{D}}{\sinh (1/\lambda )}  \label{Delta}
\end{equation}%
where $\lambda $\ is related to GBEC BF interaction parameter $f$ through%
\[
\lambda \equiv VN(0)=f^{2}N(0)/2\hbar \omega _{D}. 
\]%
This makes the first term on the rhs of (\ref{ocuarto})\ exactly equal to $%
\Delta ^{2}N(0)/\lambda $ which in turn can be shown to cancel exactly
against the log term if one recalls the hyperbolic-function identity $\sinh
^{2}x+1\equiv \cosh ^{2}x$. Thus, the GBEC theory condensation energy (\ref%
{ocuarto}) is identical for {\it any }coupling to that of BCS theory, Eq.
(2.42) of Ref. \cite{bcs}, namely 
\begin{eqnarray}
\frac{E_{s}-E_{n}}{L^{d}} &=&N(0)(\hbar \omega _{D})^{2}\left[ 1-\sqrt{%
1+\left( \Delta /\hbar \omega _{D}\right) ^{2}}\right] \hspace{1cm}%
\hbox{(BCS)}  \label{Ece} \\
&&\smash{\mathop{\relbar\joinrel\longrightarrow}\limits_{\lambda \rightarrow
0}}-\frac{1}{2}N(0)\Delta ^{2}\left[ 1-\frac{1}{4}\left( \frac{\Delta }{%
\hbar \omega _{D}}\right) ^{2}+O\left( \frac{\Delta }{\hbar \omega _{D}}%
\right) ^{4}\right] .  \nonumber
\end{eqnarray}%
This energy, associated with the expectation value of the BCS trial
wavefunction gives a rigorous upper bound to the exact ground-state energy
of the BCS Hamiltonian. Empirically, for niobium (Nb, bcc, $T_{c}\simeq 9.3K$%
, critical magnetic field $H_{c}\simeq 160kA/m$) the condensation energy to
be compared with the BCS result (\ref{Ece}) works out to be just $2\times
10^{-6}$eV/atom \cite{Annett}. The equivalence of (\ref{ocuarto}) and (\ref%
{Ece}) seems to suggest that, as in the GBEC theory, there are no pair-pair
interactions in the BCS theory either, as is evident from Hamiltonians (\ref%
{H0}), (\ref{Hint}) and (\ref{HintBCS}).

What happens on ignoring {\it either }2e- or 2h-CPs, as seems to be common
practice in theories of SCs and SFs? This gives crossover versions C and D.
Starting from (\ref{omega}) for $T=0$, and following a similar procedure to
arrive at (\ref{omega_s}) but {\it without} 2h-CPs such that $f_{-}=0$, $%
m_{0}(0)=0$ and $n_{0}(0)=\Delta ^{2}/f^{2}$, one gets 
\begin{equation}
\left[ \frac{\Omega _{s}(T=0)}{L^{d}}\right] _{+}= \hbar \omega
_{D}n_{0}(0)+2\int_{-\mu }^{0}d\xi N(\xi )\xi +N(0)\int_{0}^{\hbar \omega
_{D}}d\xi \left( \xi -\sqrt{\xi ^{2}+\Delta ^{2}}\right) .  \label{omega_s+}
\end{equation}%
Subtracting (\ref{omega_n}) from (\ref{omega_s+}) gives 
\begin{equation}
\left[ \frac{E_{s}-E_{n}}{L^{d}}\right] _{+}=\hbar \omega
_{D}n_{0}(0)+N(0)\int_{0}^{\hbar \omega _{D}}d\xi \left( \xi -\sqrt{\xi
^{2}+\Delta ^{2}}\right)  \label{condener+}
\end{equation}%
which is just {\it half }the full GBEC theory\ result (\ref{ocuarto}).
Furthermore, if $[(E_{s}-E_{n})/L^{d}]_{-}$ is the contribution from 2h-CPs
alone we may assume that $f_{+}=0$ and $n_{0}(0)=0$\ and eventually arrive
at precisely the rhs of (\ref{condener+}) but with $m_{0}(0)=\Delta
^{2}/f^{2}$\ in place of $n_{0}(0)=\Delta ^{2}/f^{2}$. Hence 
\begin{eqnarray}
\left[ \frac{E_{s}-E_{n}}{L^{d}}\right] _{+} &=&\left[ \frac{E_{s}-E_{n}}{%
L^{d}}\right] _{-}=\frac{1}{2}N(0)(\hbar \omega _{D})^{2}\left[ 1-\sqrt{%
1+\left( \Delta /\hbar \omega _{D}\right) ^{2}}\right]  \nonumber \\
&&\smash {\mathop{\relbar\joinrel\longrightarrow}\limits_{\lambda \to 0}}-%
\frac{1}{4}N(0)\Delta ^{2}\left[ 1-\frac{1}{4}\left( \frac{\Delta }{\hbar
\omega _{D}}\right) ^{2}+O\left( \frac{\Delta }{\hbar \omega _{D}}\right)
^{4}\right]
\end{eqnarray}%
which again is just one half the full GBEC theory condensation energy (\ref%
{ocuarto}) that was found to be identical to the full BCS condensation
energy (\ref{Ece}). Though not too surprising as the function $E(\epsilon
)\equiv \sqrt{(\epsilon -\mu )^{2}+\Delta ^{2}(\epsilon )},$ where $\Delta
(\epsilon )\equiv \sqrt{n_{0}}f_{+}(\epsilon )+\sqrt{m_{0}}f_{-}(\epsilon ),$
becomes ``half-gapless'' in either crossover versions C or D, this one-half
difference occurs precisely because either $n_{0}$ or $m_{0}$, {\it and }$%
f_{+}(\epsilon )$ or $f_{-}(\epsilon ),$ have been deleted. Including {\it %
both} 2e- and 2h-CPs gave similarly striking conclusions on generalizing via
the Bethe-Salpeter equation the ordinary \cite{Coo} CP problem from
unrealistic infinite-lifetime pairs to the physically expected
finite-lifetime ones of Refs. \cite{Honolulu,ANFdeLl}.\ 

\section{Conclusions}

The recent generalized BEC (GBEC) statistical single-band theory was
employed to distinguish four different versions of the BCS-Bose crossover
picture. One of these is the original BCS-Bose crossover theory with number
equation (\ref{BCSnumbereqn}), crossover version B. For the Cooper/BCS model
interaction predicts in 2D virtually the same $T_{c}$s to well beyond
physically unreasonable values of coupling, as the (allegedly less general)
BCS statistical theory where the number equation becomes trivial on assuming
that the electron chemical potential $\mu =E_{F}$, the Fermi energy.
However, $T_{c}$s much higher than those of the BCS-Bose crossover theory
have been obtained \cite{AIP}\ via the GBEC number equation (\ref{57'}),
designated here as crossover version A, that includes both electron- or
hole-pair bosons explicitly but in {\it different} proportions.

The GBEC statistical theory also reveals that the BCS gap equation for all
temperatures follows rigorously only when {\it neither} hole- nor
electron-pairs are ignored and occur in {\it equal} proportions, separately
for zero- and nonzero-CMM pairs, and that the resulting GBEC $T=0$
condensation energy equals the entire (rigorous-upper-bound) BCS value for 
{\it any }coupling. But that it is only {\it half as large }when either kind
of pair is ignored. Hence, if a BEC theory that reduces properly to BCS
theory is at all relevant in SCs and SFs taken as many-fermion systems where
pairing into bosonic CPs can occur, two-hole CPs must play an unambiguously
crucial role.

\bigskip

{\bf Acknowledgements }MdeLl thanks J.F. Annett, D.M. Eagles, M. Fortes, J.
Javanainen, T.A. Mamedov, J.P. Rodr\'{\i}guez, O. Rojo, V.V. Tolmachev, J.A.
Wilson and G.M. Zhao for discussions. MdeLl and MAS acknowledge
UNAM-DGAPA-PAPIIT (Mexico) for grants IN106401 \& IN111405, and CONACyT
(Mexico) grants 41302F \& 43234F, for partial support. MdeLl also thanks the
Texas Center for Superconductivity, University of Houston, Houston, TX 77204
for partial support, as well as the Royal Society (UK) and the Academia
Mexicana de Ciencias (Mexico) for a joint fellowship.

\end{document}